\documentclass[%
reprint,
 amsmath,amssymb,
 aps,
 pra,
]{revtex4-1}

\usepackage{graphicx}
\usepackage{dcolumn}
\usepackage{bm}
\usepackage{xcolor}
\usepackage{hyperref}
\hypersetup{colorlinks = true,allcolors = {blue}}

\begin{document}
\title{What Makes a Good Descriptor for Heterogeneous Ice Nucleation on OH-Patterned Surfaces}

\author{Philipp Pedevilla}
\affiliation{Thomas Young Centre and London Centre for Nanotechnology, 17-19 Gordon Street, London, WC1H 0AH, United Kingdom}
\affiliation{Department of Chemistry, University College London, 20 Gordon Street, London, WC1H 0AJ, United Kingdom}

\author{Martin Fitzner}
\affiliation{Thomas Young Centre and London Centre for Nanotechnology, 17-19 Gordon Street, London, WC1H 0AH, United Kingdom}
\affiliation{Department of Physics and Astronomy, University College London, Gower Street, London, WC1E 6BT, United Kingdom}

\author{Angelos Michaelides}
 \email{angelos.michaelides@ucl.ac.uk}
\affiliation{Thomas Young Centre and London Centre for Nanotechnology, 17-19 Gordon Street, London, WC1H 0AH, United Kingdom}
\affiliation{Department of Physics and Astronomy, University College London, Gower Street, London, WC1E 6BT, United Kingdom}

\date{\today}

\begin{abstract}
Freezing of water is arguably one of the most common phase transitions on Earth and almost always happens heterogeneously. Despite its importance, we lack a fundamental understanding of what makes substrates efficient ice nucleators. Here we address this by computing the ice nucleation (IN) ability of numerous model hydroxylated substrates with diverse surface hydroxyl (OH) group arrangements. Overall, for the substrates considered, we find that neither the symmetry of the OH patterns nor the similarity between a substrate and ice correlate well with the IN ability. Instead, we find that the OH density and the substrate-water interaction strength are useful descriptors of a material's IN ability. This insight allows the rationalization of ice nucleation ability across a wide range of materials, and can aid the search and design of novel potent ice nucleators in the future. 
\end{abstract}

\maketitle

\section{Introduction}
Nucleation is a process that plays a pivotal role in numerous fields. Self-assembly during biomineralisation~\cite{He_Natmat_2003_ApatiteSelfAssembly} or nanostructure formation~\cite{Jonkheijm_Science_2006_SolvAssistedNucleation}, epitaxial growth of semiconductor heterostructures~\cite{Chen_1996_PRL_StructTransHeteroepitaxy} or the controlled formation of quantum dots through heteroepitaxy~\cite{Yang_PRL_2004_QDHeteroepitaxy} are just some examples. One of the most common nucleation processes on Earth is the freezing of water, and despite its ubiquity, pure water is surprisingly difficult to freeze. Although frost on car wind shields and ice accumulation in the freezer compartment of a fridge are common annoyances that suggest otherwise, the ease of water freezing on its own can hardly be blamed for such events. The thermodynamic freezing point of water is 0 $^{\circ}$C, pure water however can remain in its liquid state until -43 $^{\circ}$C~\cite{Sellberg_Nature_2014_Xray-droplets-homo}. Indeed, most ice on Earth does not freeze by itself (homogeneously), but instead with the help of a large variety of different substrates (heterogeneously). These substrates can be mineral dust, soot particles, organic and even biological materials~\cite{Murray_ChemSocRev_2012_IN-in-clouds}. It will not come as a surprise, that the presence of such particles in, for example, clouds plays a crucial role in determining the amount of ice in them, which in turn has implications for Earth's climate~\cite{Pratt_NatGeosci_2009_bio-in-clouds, Murray_ChemSocRev_2012_IN-in-clouds, Bartels_Nature_2013_ice-chemistry, Murray_Science_2017_PerspectiveKFeldspar}.
\begin{figure*}[ht]
\includegraphics[width=16cm]{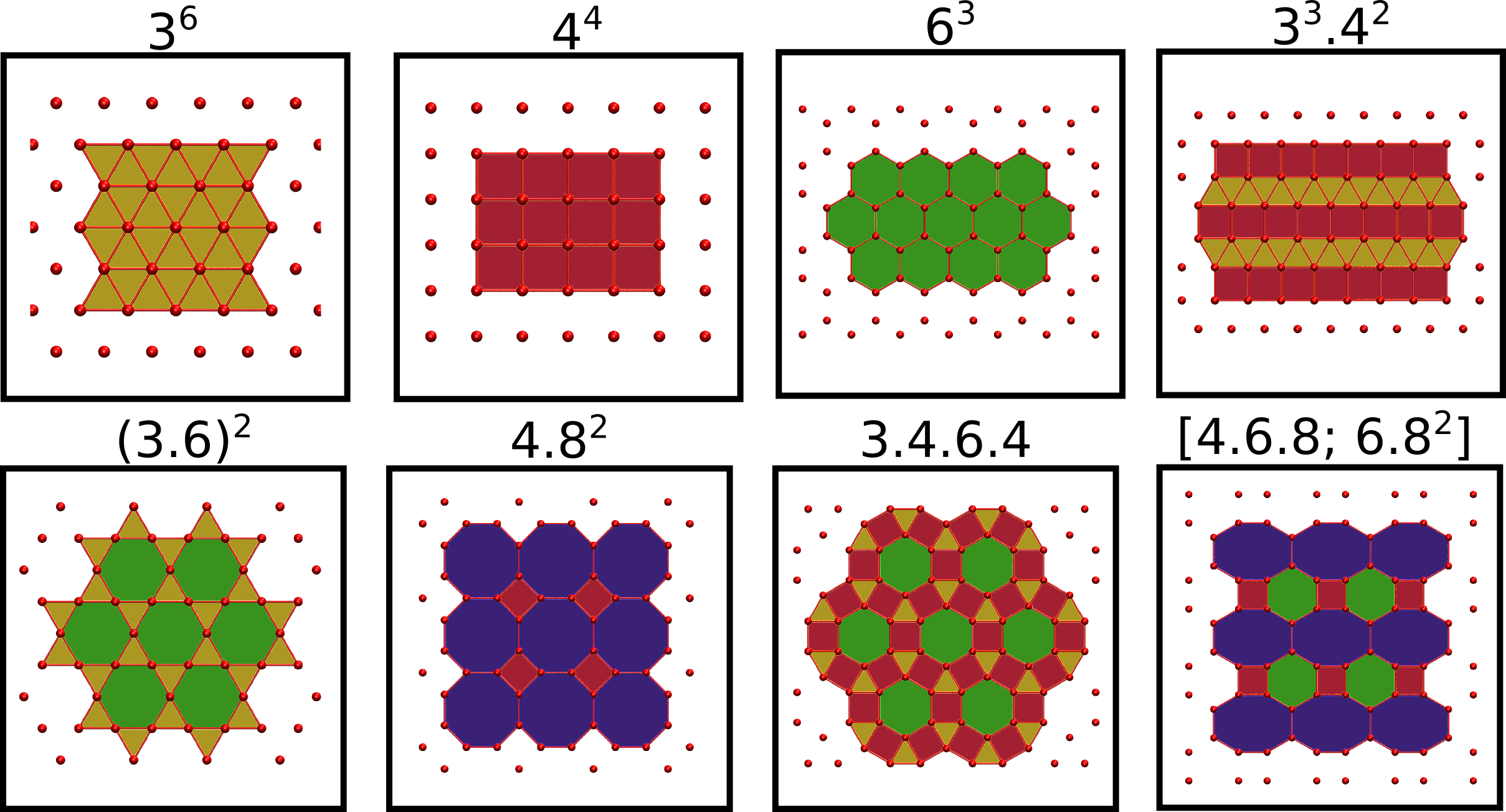}
\protect\caption{Range of OH tiling patterns considered in this study with standard vertex notation on top of each structure. All OH patterns consist of triangles (yellow), squares (red), hexagons (green), octagons (blue) and various combinations thereof.}
\label{fig: structures}
\end{figure*}

A considerable body of experimental work has been carried out with a view to understanding ice formation. Surface science measurements have provided an atomistic understanding of the initial stages of water clustering and ice formation on well-defined atomically flat surfaces~\cite{Carrasco_NatMat_2009_pentagons, Forster_2011_PRL_CuOHIceGrowth}, but these experiments are currently not applicable under atmospherically relevant conditions. Scanning electron microscopy (SEM) measurements provide insight into nucleation at a micron level, thus shedding light on the dependence of ice nucleation ability on surface topology. Through such experiments it was, for example, revealed that edges and cracks at mineral surfaces can play a crucial role in nucleating ice~\cite{Kiselev_Science_2016_Fsp100Nuc, Wang_PCCP_2016_KaoEdgeNucleation}. Droplet freezing experiments have made the systematic screening of numerous materials' ice nucleating ability possible~\cite{Atkinson_Nature_2013_fsp-IN}. This has revealed insight that e.g. oxidizing graphene flakes improves their nucleating ability~\cite{Whale_2015_JPCL_OxidizedNanomaterials}. However, neither SEM nor droplet freezing measurements provide direct molecular-level information about \emph{how} and ultimately \emph{why} a substrate is able to aid the formation of ice. What is currently missing, simply stated, is an understanding of what it is that makes materials good or bad ice nucleators. Having this knowledge would have wide-ranging implications in areas such as cryotherapy, aviation, the oil industry and the atmospheric sciences.

One potential way to tackle this issue is with computer simulations, and studying nucleation \emph{in silico} is indeed a thriving field (see e.g.~\cite{PhysRevLett.92.040801, PhysRevLett.94.235703, PhysRevLett.100.020603, Sosso_ChemRev_2016_IceReview, Zielke_JPCB_2015_kaolinite-IN, Akbari_2015_PNAS_FFS-IN-homo, Lupi_JPCA_2013_Hydrophilicity, Lupi_JACS_2014_Ccurvessize, Qiu_JACS_2017_OHnucleation, Zhang_JCP_2014_surface-structure-INA, Reinhardt_JCP_2014_surface-INA,  Cox_JCP_2015_layers, Cox_JCP_2015_hydrophilicity, Fitzner_JACS_2015_HeteroNuc-LJ,bi2017enhanced}). For ice nucleation in particular it became feasible only recently to use all-atom force field models~\cite{Espinosa_JChemPhys_2016_Homo-Seeding, Akbari_2015_PNAS_FFS-IN-homo, Zielke_JPCB_2015_kaolinite-IN, Sosso_JPCL_2016_FFS-kaolinite,Akbari_2017_PNAS_SURFACE_FREEZING}. However, looking at a wide variety of different substrates in order to extract general trends with such force fields is still out of reach. Advances in force field representations of water, such as the mW potential~\cite{Molinero_JPCB_2008_mW} have made it feasible over the past few years to study a variety of model surfaces and their impact on IN ability~\cite{Moore_PhysChemChemPhys_2010_nanopore, Lupi_JACS_2014_Ccurvessize, Zhang_JCP_2014_surface-structure-INA, Reinhardt_JCP_2014_surface-INA,  Cox_JCP_2015_layers, Cox_JCP_2015_hydrophilicity, Fitzner_JACS_2015_HeteroNuc-LJ, Bi_JPhysChemC_2016_CrystallinityHydrophilicity, Lupo_JChemPhys_2016_interfacialwater-twostep}. Despite giving valuable insight into heterogeneous ice nucleation, a clear picture of what makes a good ice nucleator remains elusive.

Here we focus on this issue, by providing an interpretation which is able to explain and therefore predict ice nucleation ability on hydroxylated model surfaces with high accuracy. The surfaces of many good ice nucleators, ranging from inorganic to organic and even biological material, are hydroxylated. Therefore understanding the connection between OH group arrangement and ice nucleation efficiency is of paramount importance. Whereas some ice nucleators, such as kaolinite, have OH groups arranged in a structure that resembles ice~\cite{Hu_SurfSci_2008_kaoliniteDFT}, other potent ice nucleators such as feldspar~\cite{Atkinson_Nature_2013_fsp-IN} and cholesterol~\cite{Ewing_ChemRev_2006_testosterone} have no apparent lattice match with ice. Furthermore, well-defined surface science measurements of ice growth on Cu have revealed that free OH groups indeed act as nucleation sites for water adsorption and subsequent ice growth~\cite{Forster_2011_PRL_CuOHIceGrowth}. With this in mind we computed the nucleation rate constants of a large variety of structurally diverse model hydroxylated substrates, all with distinct OH group patterns. We find that, simply put, if adsorption of small water clusters is weak, water molecules can rearrange in the contact layer, and ice can form, even if the substrate does not resemble an ice-like structure. Conversely, if adsorption of such clusters is strong, this rearrangement cannot take place, and ice formation is not promoted. Calculating the adsorption properties of small water clusters with, for example, all-atom force fields or \emph{ab initio} calculations can be readily done nowadays, thus this work opens up the possibility of fast and efficient predictions of the ice nucleating potential of substrates in general.

The rest of this paper is structured as follows: In section \ref{sec: CompMeth} we describe how we constructed the model substrates and the computational approach used to obtain nucleation rates. This is followed in section \ref{sec: Results} with analysis and discussion on descriptors used to characterize IN ability. We then follow this in section \ref{sec: Discussion_Conclusion} with a brief discussion of how these descriptors can be applied to some realistic materials and surfaces along with a summary of our results.
\begin{figure*}
\includegraphics[width=16.4cm]{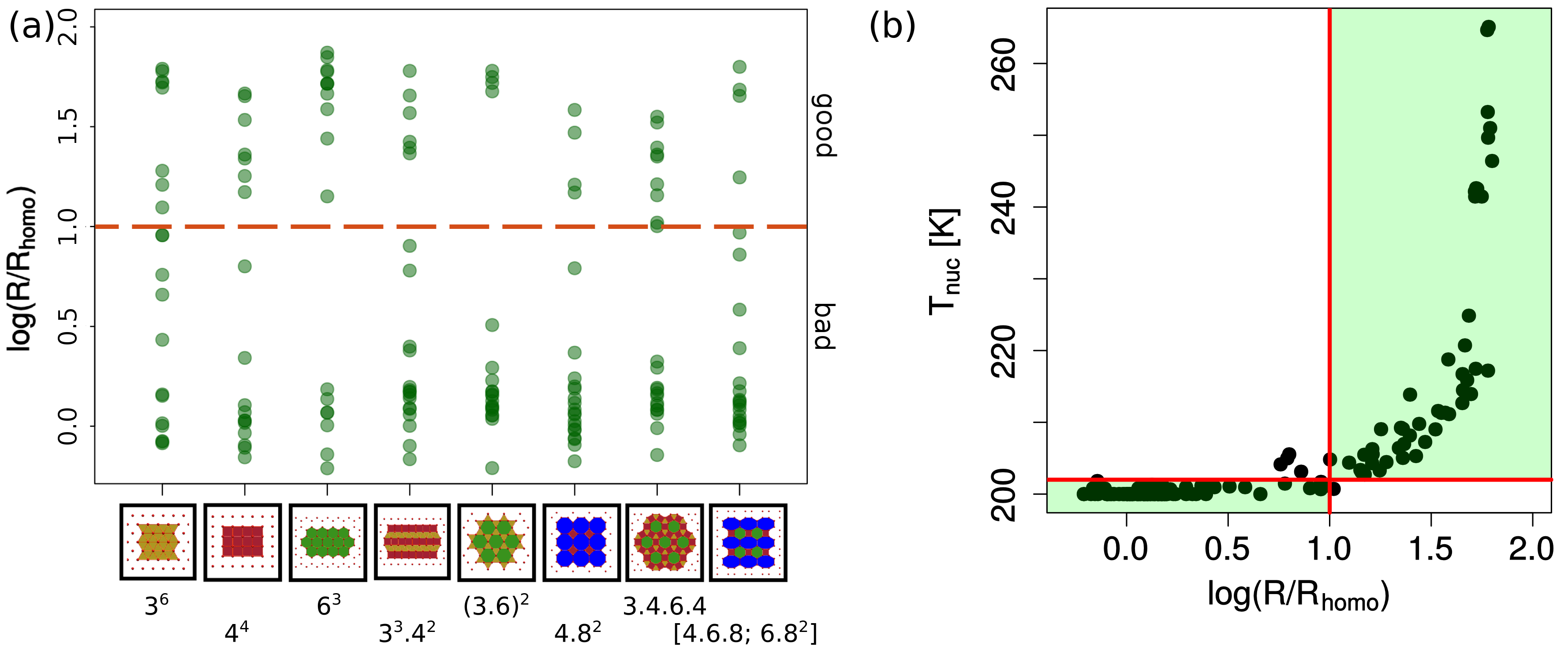}
\caption{(a) Ice nucleation rate constants relative to the homogeneous nucleation rate constants for the range of OH tiling patterns considered. Independent of the symmetry of the substrate, both good and bad ice nucleation efficiencies exist. The color code used to label polygons is the same as in figure~\ref{fig: structures}. Good and bad ice nucleators are separated by the dashed line, with good IN being at least 10 times more efficient compared to homogeneous nucleation. (b) Classification of IN ability depending on the computational approach to induce nucleation. Nucleation rate constants are plotted on the x axis, and freezing temperatures on the y axis. The red lines show the boundary between good and bad ice nucleators for both approaches.}
\label{fig: nuc-rates}
\end{figure*}

\section{Computational Approach} \label{sec: CompMeth}
We considered eight different tilings of the plane based on triangles, squares, hexagons, octagons and combinations thereof, see figure~\ref{fig: structures}. These particular polygons were chosen, because hydroxyl groups and water molecules in realistic systems are most likely to form structures based on these shapes. On kaolinite, for example, hydroxyl groups are arranged in a triangular pattern~\cite{Hu_SurfSci_2008_kaoliniteDFT}. Square-like arrangements of water molecules have also been suggested~\cite{Algara_Nature_2015_square-ice, Chen_PhysRevLett_2016_2Dice}. On some cholesterol faces, hydroxyl groups are also arranged in a nearly square-like fashion~\cite{Craven_1976_Nature_Cholesterol-MH}. Hexagonal arrangements of water molecules not only make up ice I, but can also be found on for example microcline (001)~\cite{Pedevilla_JPhysChemC_2016_feldspar-DFT}. Also unconventional water structures consisting of a mix of octagons and squares are known~\cite{Yang_PhysRevLett_2004_ice-oct-squ-tess}, and hence have been considered in this study as well. For each tiling, we varied the nearest neighbor OH distance (d$_\text{OH-OH}$) between 2 \AA~(high density OH patterns) and 6 \AA~(relatively low density OH patterns). This range of OH separations covers the range of OH separations for existing materials, and with future experiments using controlled hydroxylation of inorganic substrates in mind~\cite{Yang_JPhysChemC_2013_OHMetalSuppSilica, Yang_JPhysChemLett_2014_TuningSpatialOHArr, Yu_PCCP_2016_ElectronStimulatedHydroxylation}, extends beyond this range in both the low and high density regimes. It is not, however, designed to encompass substrates that are only very sparsely decorated in OH groups such as the predominantly graphitic surfaces examined in~\cite{Lupi_JPCA_2013_Hydrophilicity}.

We used the mW model~\cite{Molinero_JPCB_2008_mW} to represent water-water interactions. The mW water model has proven very successful in studying the behavior of water, and in particular ice nucleation~\cite{Moore_PhysChemChemPhys_2010_nanopore, Lupi_JACS_2014_Ccurvessize, Zhang_JCP_2014_surface-structure-INA, Reinhardt_JCP_2014_surface-INA,  Cox_JCP_2015_layers, Cox_JCP_2015_hydrophilicity, Fitzner_JACS_2015_HeteroNuc-LJ, Bi_JPhysChemC_2016_CrystallinityHydrophilicity, Lupo_JChemPhys_2016_interfacialwater-twostep}. The mW potential for example describes the density of both water and ice better than commonly used all-atom force fields, and also describes the nucleation of stacking disordered ice well. The interested reader is referred to the supporting information (SI~\cite{note_SI2}), where we discuss in more detail the suitability of this potential for studying aspects of ice nucleation ability. Hydroxyl groups were modeled as frozen mW molecules, as was done before~\cite{Lupi_JPCA_2013_Hydrophilicity}. Note that although the positions of the OH groups are frozen, the hydrogen bond directions of the OH groups are not. This is because the hydrogen bond network arrangement is simulated by means of a three body potential, which can act in any direction. Therefore, the hydrogen bond network that the OH groups induce is fully flexible. For each OH pattern, two model substrate structures were obtained by adding a Lennard-Jones (9,3) wall-potential in the z direction at the bottom of a simulation cell (with interaction strength of either $\varepsilon=0.05$ eV or $\varepsilon=0.20$ eV). The resulting adsorption energies cover a physically reasonable range, from -0.11 eV (weak adsorption as found on hydrophobic surfaces) to -0.88 eV (strong adsorption as found on hydrophilic surfaces) per water molecule. Note that we did not tune the adsorption energies to fall within this range, instead they emerged naturally from the large structural space of OH arrangements. We therefore believe that the substrate space considered in this work mimics a broad range of realistic systems, without representing any particular system explicitly. 4000 water molecules were placed on top of each surface, which is sufficient to restore bulk densities of liquid water~\cite{Fitzner_JACS_2015_HeteroNuc-LJ}. The lateral dimensions of the simulation boxes were chosen to be as close as possible to 50 \AA~$\times$ 50 \AA, and never less than 40 \AA~$\times$ 40 \AA. A total of 164 different substrates were generated in this way. 
\begin{figure*}
\includegraphics[width=16.0cm]{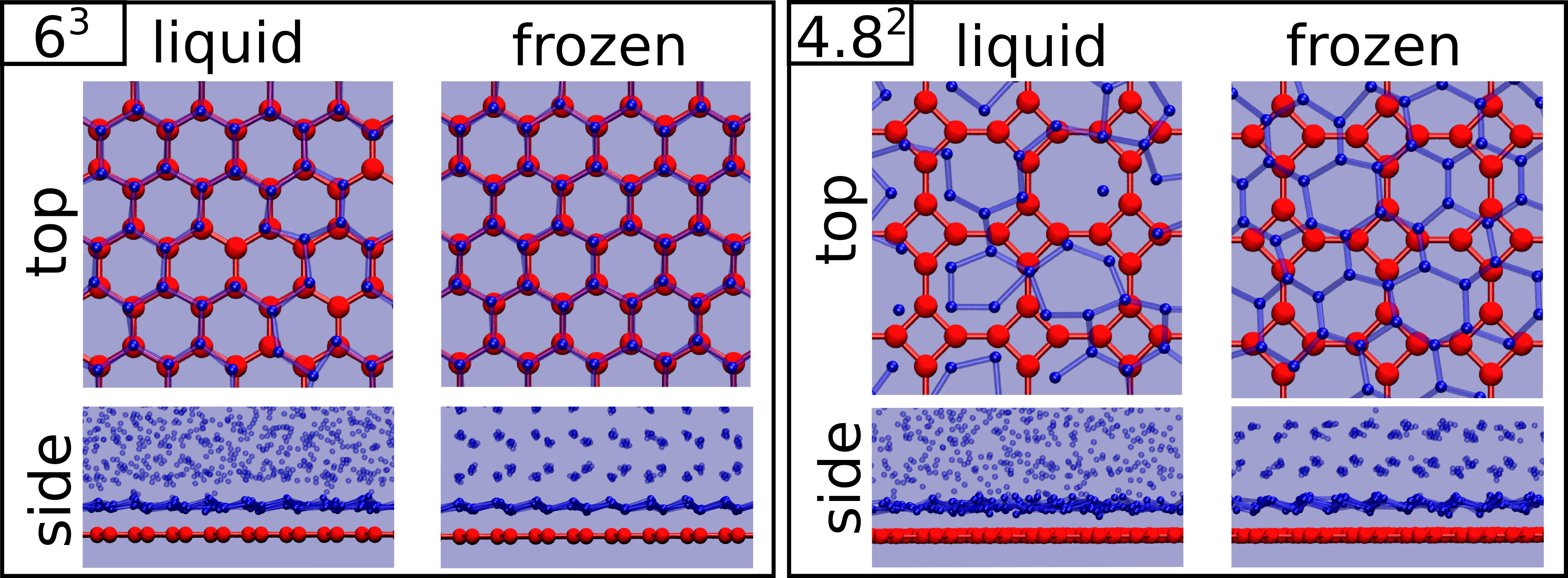}
\caption{Contact layer rearrangement upon freezing. The figure shows the water interface before and after ice nucleation for a substrate that matches the structure of ice (6$^3$), and one that does not (4.8$^2$). OH groups are shown in red, and water molecules in blue. In the top-view images, only the water molecules in the contact layer are displayed.}
\label{fig: interface}
\end{figure*}

Nucleation rate constants were computed by means of brute force molecular dynamics simulations at 205 K with an established protocol~\cite{Cox_JCP_2015_layers, Cox_JCP_2015_hydrophilicity, Fitzner_JACS_2015_HeteroNuc-LJ}. In brief, the procedure involved quenching 15 simulations to the target temperature and measuring the induction times for nucleation from which we estimated the nucleation rate. The induction times were identified by the drop in potential energy which coincides with the nucleation of sizeable ice-clusters (see e.g.~\cite{Fitzner_JACS_2015_HeteroNuc-LJ}). Further details can be found in the SI~\cite{note_SI2}. The average freezing temperature obtained from progressively decreasing the temperature of 5 simulations for each substrate by $-1$~K/ns starting at 270~K was also calculated, more details about this procedure can be found in references~\citenum{Lupi_JPCA_2013_Hydrophilicity, Lupi_JACS_2014_Ccurvessize, Qiu_JACS_2017_OHnucleation}.

\section{Results} \label{sec: Results}
Figure~\ref{fig: nuc-rates}(a) shows the nucleation rate relative to the homogeneous one as a function of the underlying substrate pattern. The combination of supercooling and simulation time (hundreds of ns for the longest) in our study leads to nucleation rates within a few orders of magnitude. A nucleation rate that is two orders of magnitude larger than the homogeneous rate is the largest rate achieved, similar to previous work~\cite{Fitzner_JACS_2015_HeteroNuc-LJ}, and corresponds to immediate freezing within the first few nanoseconds. We therefore focus on the more qualitative distinction between good and bad nucleators, and define good ice nucleators to be at least 10 times more efficient than homogeneous nucleation, and label everything that nucleates ice at a slower rate than that a bad nucleator. Note that the distinction between good and bad ice nucleators is a binary classification aimed at finding the key differences between substrates that are able to accelerate ice nucleation and ones which do not. Figure~\ref{fig: nuc-rates}(b) shows the correlation between nucleation rate constants and freezing temperatures. Apart from the strong correlation between the two methods, the classification of good and bad ice nucleators also agrees well between the two approaches. In the case of cooling ramps, we define a good ice nucleator to nucleate above the homogeneous freezing temperature of mW water at a cooling rate of $-1$ K/ns. The red lines in figure~\ref{fig: nuc-rates}(b) show the boundaries between good and bad ice nucleators, and the areas in green highlight the regimes in which both protocols lead to the same classification of IN ability. Clearly, the vast majority of points is assigned in the same fashion, irrespective of the computational protocol used to calculate IN ability. Furthermore, the sparsity of the nucleation rates around the boundary between good and bad ice nucleators combined with similar results obtained via two different approaches means that our results are not dependent on the details of the choice of this classification boundary (see SI~\cite{note_SI2}).

From the data reported in figure~\ref{fig: nuc-rates}(a) it is clear that for every OH pattern good and bad nucleating efficiencies can be found. Hexagonal surfaces ($6^3$), i.e. surfaces that resemble the symmetry of ice, are by no means better than, for example, structures that do not seem to resemble ice at all. In scenarios where the substrate structure matches the structure of ice very well ($3^6$ and $6^3$, d$_{\text{OH-OH}} \approx $ 2.7 \AA), a hexagonal contact layer akin to ice forms which leads to fast nucleation, see figure~\ref{fig: interface}. This behavior of a substrate templating the structure of ice and subsequently leading to high IN ability is not surprising. However, many other examples of fast nucleation exist, where no such templating effect is present. In these cases, the contact layer does not resemble the structure of ice, but rearranges during nucleation to form an ice-like structure (figure~\ref{fig: interface}). Independent of nucleation via templating or contact layer rearrangement, we observe stacking disordered ice to form, which is expected at strong supercooling~\cite{Malkin_PNAS_2012_StackingDisorderedIce, Kuhs_PNAS_2012_StackingDisorderedIce, Malkin_PCCP_2015_StackingDisoreredIce, Hudait_JACS_2016_IcePolymorphClouds}. We note in passing that in figure~\ref{fig: nuc-rates}(a) nucleation rate constants below the homogeneous one are reported. These do not mean that a substrate is truly inhibiting ice nucleation. The finite simulation box size rather means that a contact layer on such a substrate is deactivated and will therefore not contribute to the homogeneous ice nucleation rate constant.
\begin{figure}
\includegraphics[width=8.8cm]{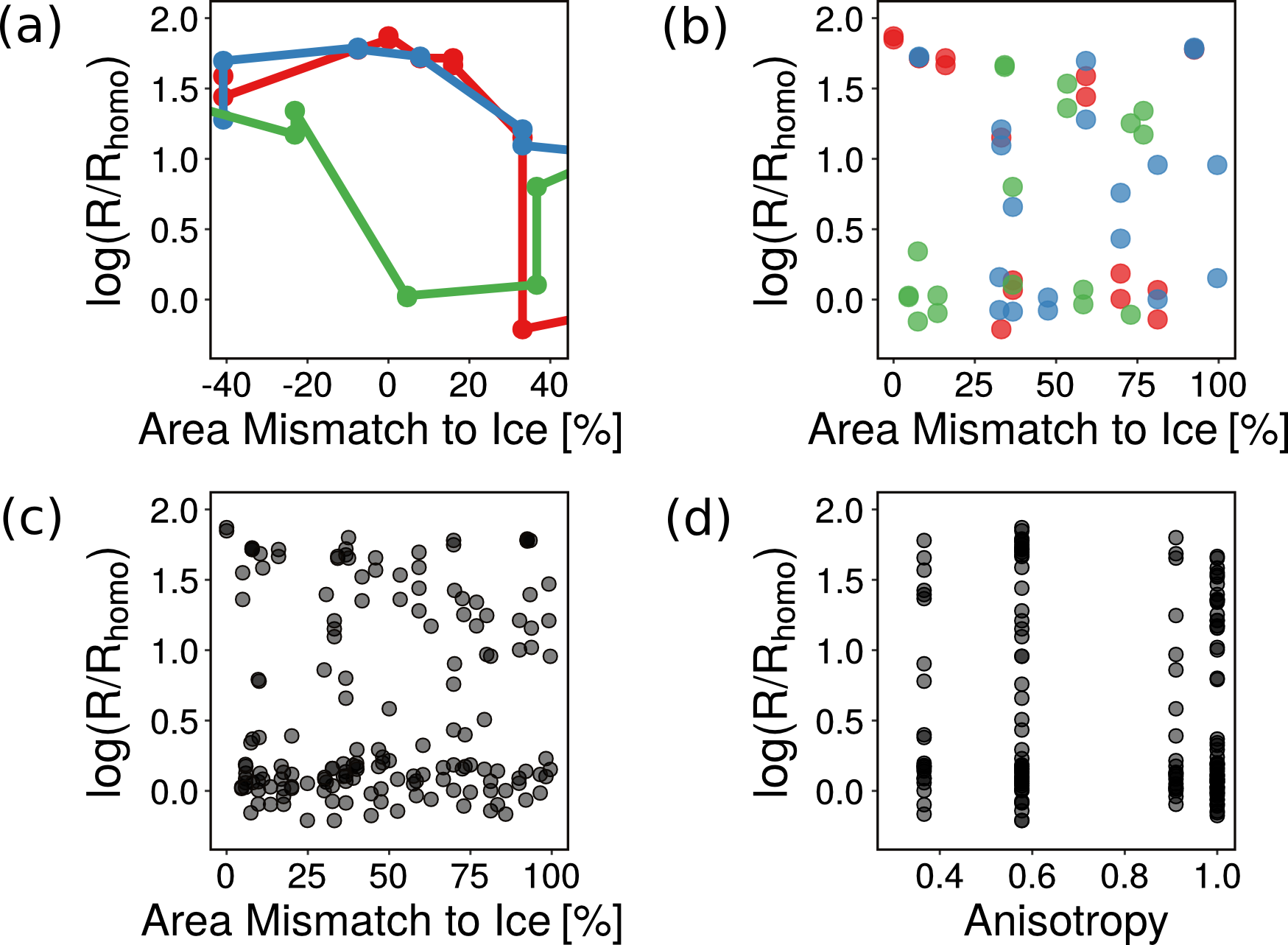}
\protect\caption{Capability of area mismatch and lattice anisotropy to describe IN ability as suggested by Qiu \emph{et al.}~\cite{Qiu_JACS_2017_OHnucleation}. (a) Our results agree with Qiu \emph{et al.}~\cite{Qiu_JACS_2017_OHnucleation}, in that a subset of structures (hexagonal, red and triangular, blue) can be described well using area match, whereas others (such as a square arrangement of OH groups, green) fall outside the scope of this descriptor. (b) The IN ability of the same structures described in (a) does not follow the expected behavior once larger deviations of lattice match are considered. (c-d) The full set of substrates considered in this study as a function of area mismatch and anisotropy respectively.}
\label{fig: mismatch}
\end{figure}

The result that the symmetry of the OH group arrangement of the substrate does not play a crucial role in determining the ice nucleation efficiency is a key finding of this work. Given the widely held view that hexagonal substrates make the best templates for ice nucleation it comes somewhat as a surprise. In the rest of this paper, we will focus on trying to explain this behavior, by identifying features that are able to discriminate between good and bad ice nucleating agents. Various descriptors were considered and we now discuss how some of the most interesting structural and energetic descriptors perform in discriminating between good and bad IN ability. 

Lattice match is frequently used to explain IN ability~\cite{Pruppacher_BOOK_1998_Microphys-Clouds}. The basic idea behind this is simple: ice can form more readily on substrates that themselves look ice-like. Recently, whilst studying the IN ability of alcohol films, Qiu \emph{et al.}~\cite{Qiu_JACS_2017_OHnucleation} showed that the area match between ice and a substrate as well as anisotropy can predict IN ability. Area match is defined as $((a_\mathrm{OH} \times b_\mathrm{OH}) / (a_\mathrm{ice} \times b_\mathrm{ice}) -1) \times 100$, and anisotropy as $b_\mathrm{OH}/(\sqrt{3}a_\mathrm{OH})$~\cite{Qiu_JACS_2017_OHnucleation}, where $a_\mathrm{OH}$ and $b_\mathrm{OH}$ are the rectangular lattice parameters of the OH structure and $a_\mathrm{ice}$ and $b_\mathrm{ice}$ are the lattice parameters of ice. Figure \ref{fig: mismatch}(a) shows how area mismatch correlates with ice nucleating efficiency for a certain range (similar to the one considered by Qiu \emph{et al.}). We find that, in line with Qiu \emph{et al.}, particular substrates are most efficient with an mismatch close to zero. This is the case for hexagonal ($6^3$) and triangular ($3^6$) arrangements shown in red / blue respectively. However, square arrangements, like e.g. ($4^4$), follow a different trend, shown in green. In addition, the predictive quality of this descriptor diminishes if we consider a larger range of area mismatch, shown in figure~\ref{fig: mismatch}(b). There, we projected all area mismatches in the range [0,100] (the largest are match considered in our study was 400\%). We do the same for the other substrates in (figure~\ref{fig: mismatch}(c)) and it becomes clear that the area match is not able to discriminate the IN ability of all substrates. The same behavior is found for the anisotropy as illustrated in figure~\ref{fig: mismatch}(d). Each of our OH symmetries is associated with an anisotropy that is independent of the OH density. However, as we already saw from figure~\ref{fig: nuc-rates}(a) the same symmetry (and hence anisotropy) can make for both good and bad ice nucleators. Overall, our findings agree with those of Qiu \emph{et al.}: If the OH arrangement resembles the structure of ice very closely, area match and anisotropy describe the nucleation enhancement well. We show however that for OH symmetries that do not belong to this specific class these descriptors do not correlate in general with the IN ability.

\begin{figure*}
\includegraphics[width=17cm]{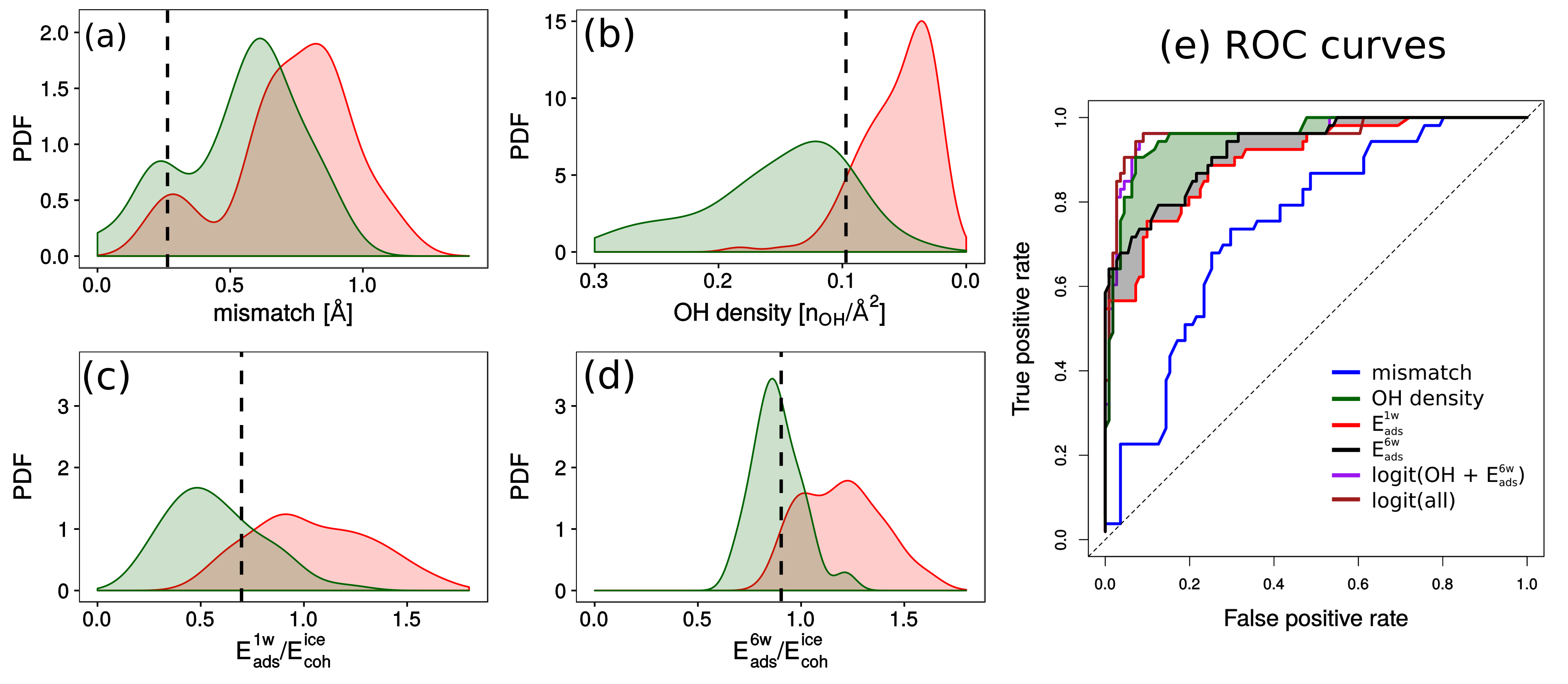}
\protect\caption{Performance of various descriptors for heterogeneous ice nucleation. Probability density function (PDF) of good (green) and bad (red) ice nucleators as a function of the mismatch between the substrates and ice (a), OH density (b), the adsorption energy of a single water ($E_\mathrm{ads}^\mathrm{1w}$) (c) and of six waters ($E_\mathrm{ads}^\mathrm{6w}$) (d). Adsorption energies are reported relative to the cohesive energy of ice ($E_\mathrm{coh}^\mathrm{ice}$). The black dashed line shows the decision boundary that corresponds to the highest classification accuracy. (e) The ROC curves for the features in (a)-(d) as well as combinations thereof in the form of logit models.}
\label{fig: descriptors}
\end{figure*}
To get a more general measure of the matching quality, we compute the match between a substrate and ice using the root mean square deviation (RMSD) of various ice crystallites on top of a surface, and report the RMSD that is associated with the ice face that matches the substrate most closely. Details of the approach are outlined in the SI~\cite{note_SI2}. For here it is sufficient to note that a small value of the mismatch signifies a good match between the substrate and ice, and a large value of the mismatch means that the substrate and ice do not match well. Figure~\ref{fig: descriptors}(a) shows the performance of this mismatch in distinguishing good and bad ice nucleators. On the x-axis, the classifier is shown and on the y-axis the probability density function (PDF) is shown for good (green) and bad (red) ice nucleators. The smaller the overlap between the two densities, the better the classifier. An ideal classifier would completely separate the two densities. It can be seen that the PDFs for the good and bad nucleating substrates overlap considerably, hence making lattice mismatch a poor descriptor in general. Clearly, therefore, the similarity between a substrate and ice is not a good way think about IN ability for the hydroxylated surfaces considered here. This is in line with previous work, e.g. simulations on LJ crystals exposing different surfaces~\cite{Fitzner_JACS_2015_HeteroNuc-LJ}.

Figure~\ref{fig: descriptors}(b) shows that OH density separates IN ability significantly better than mismatch does. This is particularly clear when comparing the overlap between good and bad nucleators in both cases. The PDF for good nucleators is close to zero when the PDF of bad nucleators reaches its maximum and vice versa. This in turn means, that a classification based on OH density will have low false positive and false negative classification rates. As a consequence, for the range of hydroxylated surfaces considered here, the OH density for regular tiling patterns is a good descriptor for IN ability: High OH densities (small d$_{\text{OH-OH}}$) and low OH densities (large d$_{\text{OH-OH}}$) correspond to good and bad IN ability respectively. This is consistent with the work of Qiu \emph{et al.}~\cite{Qiu_JACS_2017_OHnucleation} which showed that having a close area match to ice (high OH density) nucleated better than substrates with a larger mismatch. 

In previous studies, water monomer adsorption energies were shown to correlate with IN ability, however this was system dependent~\cite{Cox_JCP_2015_hydrophilicity, Fitzner_JACS_2015_HeteroNuc-LJ}. Here we find as well that the monomer can be a good descriptor. Specifically, small adsorption energies yield good nucleators whereas large adsorption energies generally lead to poor nucleating substrates. However, we also find here that adsorption energies that take into account more water molecules are consistently better. With this in mind we show in figure~\ref{fig: descriptors}(c-d) how the adsorption energy of a single water molecule and six water molecules perform in discriminating between good and bad IN ability. The PDF range of good and bad IN ability is considerably broader when using only a single water molecule compared to the case when using six water molecules, and densities between good and bad ice nucleators overlap more substantially in the former case. Moreover, the maximum of the density for bad ice nucleators overlaps more with the good ice nucleators when using only one water molecule which also makes the classification in the former case worse. The adsorption energy that most accurately discriminates between good and bad IN ability is less than the bulk cohesive energy of ice (around 90 \%, see figure~\ref{fig: descriptors}(d)). If water molecules adsorb more strongly on the surface than they would in ice, it is energetically not favorable to rearrange the contact layer into ice. If, however, adsorption is weaker, then this rearrangement of the contact layer can and does happen, and hence ice can form.

Figure~\ref{fig: descriptors}(e) shows a quantitative comparison between different classifiers (and also combinations of them) in the form of Receiver Operator Curves (ROCs). ROCs are a common way to quantify the quality of a descriptor or set of descriptors, and show the true positive rate of classification against the false positive rate. An ideal classifier would have a 100 \% true positive and 0 \% false positive classification, a completely random classifier would follow the $y=x$ line (area of 0.5), shown as dashed line in figure~\ref{fig: descriptors}(e). The performance of mismatch (blue), OH density (green), monomer adsorption energy (red) and hexamer adsorption energy (black) is shown. Additionally, we highlight the area between E$_\text{ads}^\text{1mW}$ and E$_\text{ads}^\text{6mW}$ as well as the area between E$_\text{ads}^\text{6mW}$ and the OH density to show the respective improvement. We also looked at how combinations of descriptors perform compared to single classifiers using a logistic regression model (logit model). Figure~\ref{fig: descriptors}(e) shows that the improvement from combining classifiers is only a minor one, which is in line with the fact that OH density and adsorption energy are related with each other (see SI~\cite{note_SI2}). Combining them will therefore not improve the classification substantially.

\section{Discussion and Conclusions} \label{sec: Discussion_Conclusion}
Having established a simple means of gauging the IN ability of model systems, we now show that the insights can be used to rationalize various experimental observations on hydroxylated surfaces. One of the most extensively studied systems for heterogeneous ice nucleation is the mineral kaolinite. The basal surface of kaolinite has received most attention, and is, to the best of our knowledge, the only system for which the heterogeneous ice nucleation rate has been calculated quantitatively with an all-atom force field~\cite{Sosso_JPCL_2016_FFS-kaolinite}. Interestingly water hexamers bind to kaolinite with an adsorption energy that is $\sim$90 \% the cohesive energy of ice, implying that kaolinite is an effective ice nucleating agent. In addition, the prism face nucleated on kaolinite~\cite{Cox_FaradayDiscuss_2013_MD-kao-nuc, Zielke_JPCB_2015_kaolinite-IN, Sosso_JPCL_2016_FFS-kaolinite}, which is worth pointing out because water structures found by Hu \emph{et al.} did not resemble the prism face of ice~\cite{Hu_SurfSci_2008_kaoliniteDFT}. Thus, in order for ice to form on the basal face of kaolinite, water molecules had to rearrange their structure in the contact layer. This example illustrates a key finding of this work: if the interaction between the substrate and water is not too strong, the contact layer can rearrange, in this particular case to form the prismatic face of ice.

A wide ranging implication of this work is that edges and cracks for minerals in general might be more important for IN ability than previously thought. At edges and cracks there will be more broken covalent bonds, which will be saturated with hydroxyl groups under ambient conditions, which implies that there will often be a higher OH density at defect sites. Therefore, they might play a more pronounced role than the surfaces that are exposed predominantly, and recent experimental evidence is supporting this hypothesis~\cite{Kiselev_Science_2016_Fsp100Nuc, Wang_PCCP_2016_KaoEdgeNucleation}. A recent experimental study demonstrated for example that ice does not form on the basal face of kaolinite, but instead nucleates on the edges between stacked basal platelets~\cite{Wang_PCCP_2016_KaoEdgeNucleation}. Similarly, feldspar's outstanding IN ability was attributed to the formation of ice in a defect site~\cite{Kiselev_Science_2016_Fsp100Nuc}.

To summarize, this work is a systematic study of how ice nucleation ability relates to OH patternation on hydroxylated surfaces. The striking result was that the ice nucleation ability depends less on the symmetry of the pattern itself, but rather on the OH group density. An interpretation that explains this was developed based on the interaction strength between water and the substrate relative to the bulk cohesive energy of ice. Substrates that template ice-like contact layers were found to be equally efficient ice nucleators at the supercooling in this study than non ice-like substrates, if they do not adsorb water too strongly. The rationale behind this interpretation is physically well motivated: if the surface is able to accommodate water not too strongly, the contact layer can rearrange easily to reach the basin of ice. If however water adsorbs more strongly than the energetic gain from this, the energy cost associated with rearranging the structure is too high, and ice cannot form efficiently on such a substrate. The information required to estimate IN ability is accessible experimentally and computationally, and hence we believe our interpretation shows new ways forward to understand heterogeneous ice nucleation in the future.  

\begin{acknowledgments}
This work was supported by the European Research Council under the European Union’s Seventh Framework Programme (FP/2007-2013) / ERC Grant Agreement No. 616121 (HeteroIce project). A.M. is also supported by the Royal Society through a Royal Society Wolfson Research Merit Award. We are grateful for computational resources provided by the Materials Chemistry Consortium through the EPSRC Grant No. EP/L000202, the London Centre for Nanotechnology and Research Computing at University College London.
\end{acknowledgments}

\bibliographystyle{apsrev4-1} 
\bibliography{manuscript}

\clearpage
\onecolumngrid 
\setcounter{section}{0}
\renewcommand{\thesection}{S\arabic{section}}%
\setcounter{table}{0}
\renewcommand{\thetable}{S\arabic{table}}%
\setcounter{figure}{0}
\renewcommand{\thefigure}{S\arabic{figure}}%
\section*{Supporting Information}
\raggedbottom

\subsection{Contact Layer Rearrangement}
\begin{figure}[ht]
\begin{centering}
\centerline{\includegraphics[width=1.0 \textwidth]{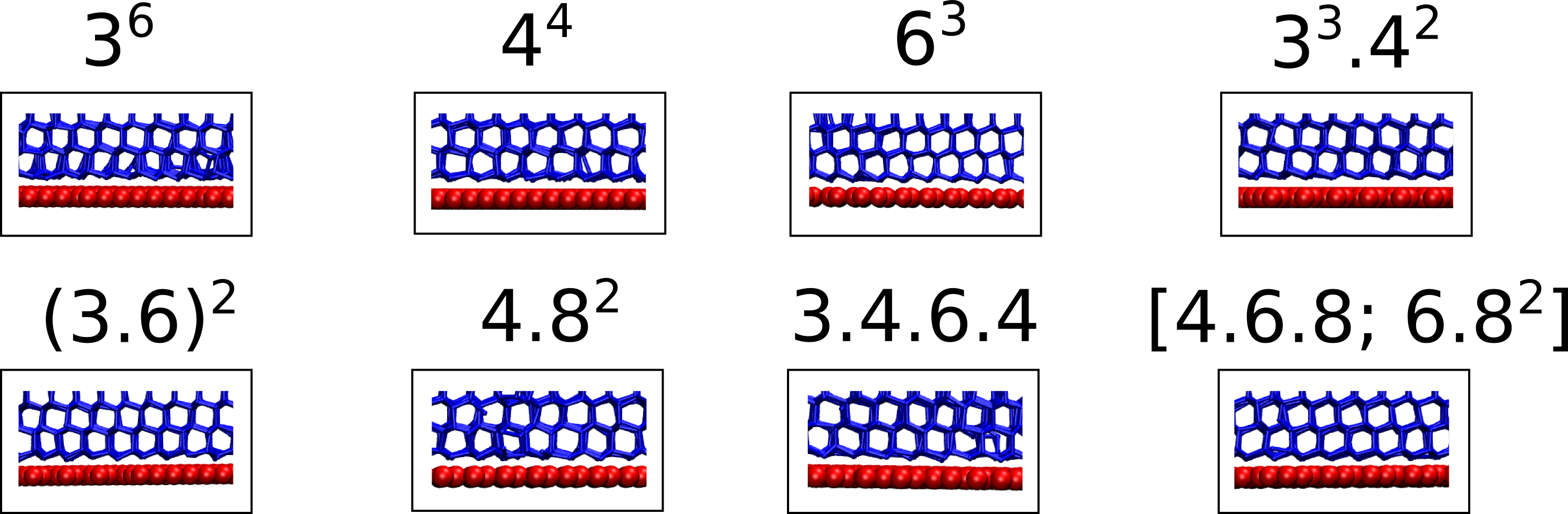}}
\par\end{centering}
\protect\caption{Simulation cells after nucleation. For all snapshots shown, d$_\mathrm{OH-OH}$ = 2.5\AA. The contact rearranges into an ice-like structure upon freezing, here resembling the basal face of ice.}            
\label{fig: CL rearrangement}
\end{figure}

In this section, we show that the contact layer does rearrange into an ice-like structure upon freezing, even when the substrate does not resemble the structure of ice. To do so, we show snapshots in figure \ref{fig: CL rearrangement} after freezing of all substrates with d$_\mathrm{OH-OH}$ = 2.5~\AA and a LJ interaction strength of 0.05 eV. This shows, that the ice formation on non ice-like substrates is not mediated through a non ice-like contact layer, and hence requires the water molecules in the contact layer to rearrange.

\clearpage
\subsection{Mismatch Calculation}
\begin{figure}[ht]
\begin{centering}
\centerline{\includegraphics[width=1.0 \textwidth]{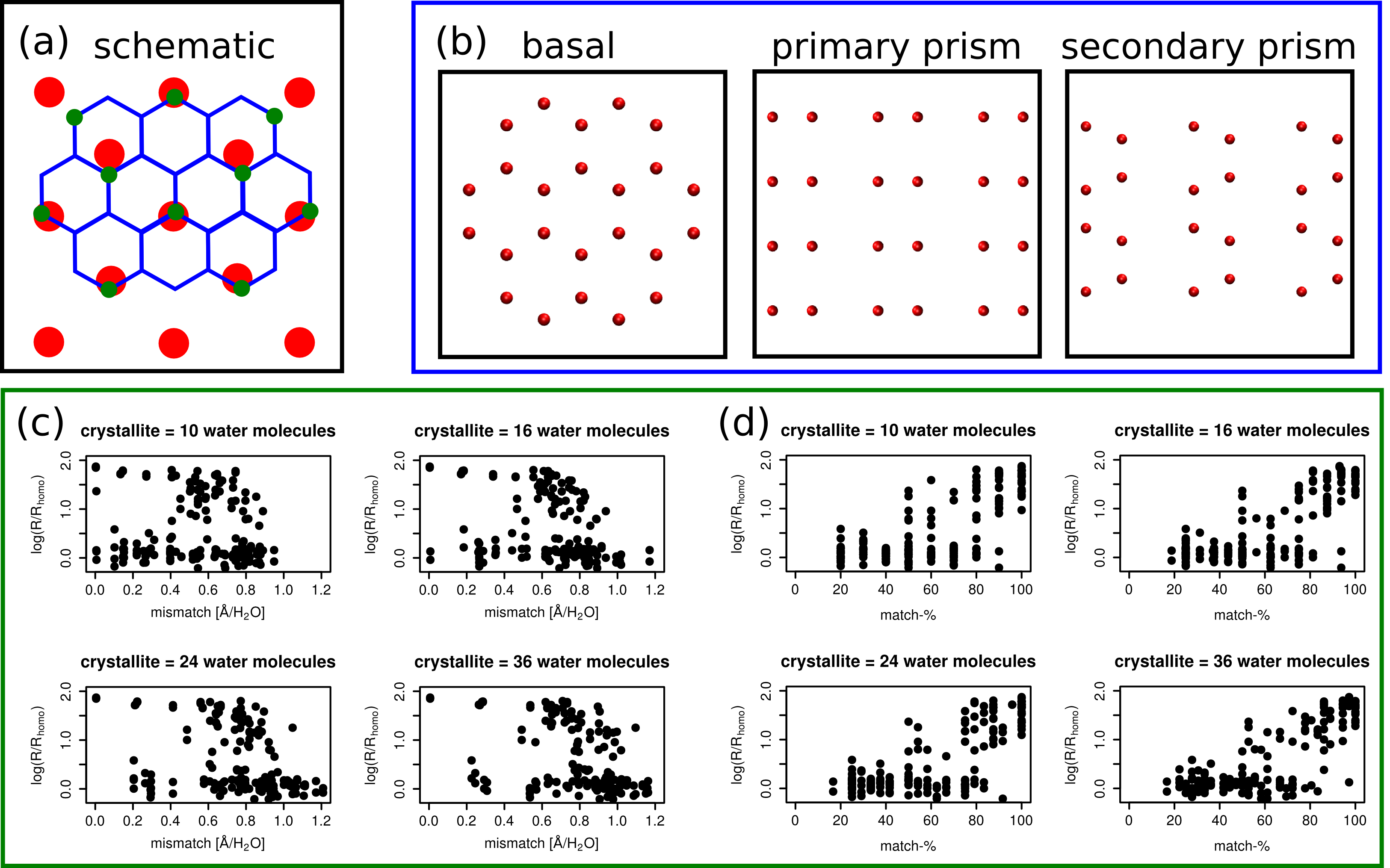}}
\par\end{centering}
\protect\caption{(a) Schematic drawing showing why not every water in a crystallite (blue) can be matched with an OH group on the surface (red). Crystallite atoms that would be used in this hypothetical example are highlighted in green, hence resulting in $N_\mathrm{M}$ = 10. (b) shows the crystallites resembling the basal, primary and secondary prism faces of ice used to calculate the mismatch, all contain 24 water molecules. (c) and (d) show how the mismatch and matching percentage change as a function of crystallite size. They are qualitatively identical, in no case do mismatch or matching percentage correlate well with INA.}            
\label{fig: mismatch2}
\end{figure}

The substrate space considered in this study is structurally diverse, in fact it is the most diverse one ever considered in a nucleation study known to the best of the authors knowledge. Because of this complexity, lattice mismatch $\delta$ traditionally defined for hexagonal surfaces~\cite{Cox_PCCP_2012_lattice-match} as
\begin{equation}
\delta = \frac{a_{0,\mathrm{s}}-a_{0,\mathrm{ice}}}{a_{0,\mathrm{ice}}}
\end{equation}
cannot be used to judge the mismatch between ice and a substrate here. Instead, we compute the smallest root mean squared deviation (RMSD) between the substrate and ice, $\zeta$.
\begin{equation}
\zeta = \min\limits_{\mathbf{r}_0, \theta} \left( \frac{\sqrt{\sum_{i=1}^{N_\mathrm{ice}} \big (\mathbf{r}_i(\mathbf{r}_0, \theta) - \mathbf{r}_\mathrm{s} \big)^2}}{N_\mathrm{M}}\right)
\end{equation}
Here, $\mathbf{r}_0$ stands for the position vector of the center of mass of the ice crystallite, $\theta$ for its rotational orientation relative to the z-axis, $N_\mathrm{ice}$ for the number of water molecules in the ice crystallite, $\mathbf{r}_i(\mathbf{r}_0, \theta)$ for the position of oxygen atom $i$ at a given $\mathbf{r}_0$ and $\theta$ and $\mathbf{r}_\mathrm{s}$ for the closest substrate OH group to water molecule $i$. If water molecule $i$ does not have a substrate atom close by (threshold: 3.3~\AA), then this water molecule is omitted for the RMSD calculation. This is physically motivated by the idea that groups that do not physically interact with each other because they are not in close proximity should not be used to judge how well ice matches with the structure of the substrate, because there is not energetic penalty for such situations. Moreover, if the number density of OH groups of the substrate is larger than the number density of water molecules in ice, it would cause fictitiously large RMSD values, if every water molecule in the crystallite would be enforced to have a nearest neighbor for the calculation of $\zeta$. $N_\mathrm{M}$ counts the number of water molecules in the crystallite that do have a substrate OH group in proximity below the threshold value, and the matching percentage is $N_\mathrm{M} / N_\mathrm{ice}$. Figure \ref{fig: mismatch2} (a) illustrates the need for $N_\mathrm{M}$ in a schematic way. 

Figure \ref{fig: mismatch2} (b) shows the ice crystallites used to match the substrate structures. The basal as well as primary and secondary prism face of ice was used to calculate $\zeta$, we always report the lowest value of these three. Because of the slight buckling within one basal face plane, all water molecules were projected into the same xy plane, and the resulting structure was used to calculate the mismatch. For the two prism faces, the buckling between hydrogen bonded water molecules within one layer is considerably larger than in the case of the basal face, hence only the low-lying water molecules have been considered for the computation of $\zeta$. All of them are larger than the underlying periodicity of the substrate.

Figure \ref{fig: mismatch2} (c) and (d) shows how the mismatch $\zeta$ and matching percentage change as a function of crystallite size respectively, only considering the basal face crystallite with size $N_\mathrm{ice}$ = {10, 16, 24, 36}. The qualitative conclusions about $\zeta$ and the matching percentage does not change with changing number of crystallite size.

\clearpage
\subsection{Computational Setup}
The method used to calculate nucleation rates in this study is well established, and the interested reader is referred to ~\cite{Cox_JCP_2015_hydrophilicity, Cox_JCP_2015_layers, Fitzner_JACS_2015_HeteroNuc-LJ} for more details. In particular, the setup employed here has been extensively tested by Fitzner et al.~\cite{Fitzner_JACS_2015_HeteroNuc-LJ}.

All MD simulations of all systems were performed with LAMMPS \cite{Plimton_JComputPhys_1995_LAMMPS}. The equations of motion for water molecules were integrated using a 10 fs timestep. Surface OH groups were kept fixed for all simulations. To calculate nucleation rate constants, the simulation box was equilibrated for 50 ns in the NVT ensemble at the beginning. Temperature was controlled with a Nos\'{e}-Hoover chain thermostat~\cite{Nose_JCP_1984_NH-thermostat, Martyna_JCP_1992_NH-chains} with chain length of 10 and set to 290 K. After equilibration, 15 initial structures separated by 5 ns each were generated under the same conditions. Each one of those 15 structures was used for other NVT simulations at 205 K to observe nucleation. Initial velocities were randomly generated according to a Gaussian distribution. The point in the trajectory where nucleation happened, the induction time $t_\mathrm{ind}$, was obtained by fitting the following equation to the potential energy $U(t)$ of the system as a function of time:
\begin{equation}
U(t)=U_{0}+\frac{\Delta U}{1+\exp(k(t-t_\mathrm{ind}))}\label{eq:U(t)}
\end{equation}
$U_{0}$, $\Delta U$, $k$ and $t_\mathrm{ind}$ are freely variable parameters. Finally a rate was calculated by fitting the decay 
\begin{equation}
P_\mathrm{liq}(t)=\exp[-(Rt)^{\alpha}]\label{eq:rate}
\end{equation}
to the probability distribution of induction times $t_\mathrm{ind}$ from the 15 simulations obtained by fitting equation \ref{eq:U(t)}. $R$ and $\alpha$ are free parameters in equation \ref{eq:rate}, where $R$ stands for the nucleation rate constant. Note that $\alpha$ is not always equal to unity, because some nucleation events happen quickly, and are actually not nucleation events in the true sense of the meaning, but rather a relaxation process. The same behavior was found by for example~\cite{Fitzner_JACS_2015_HeteroNuc-LJ}. To make the homogeneous and heterogeneous rates as comparable as possible, the rate constant $R$ was obtained per number of water molecules (which was the same for all simulations) and unit time.~\cite{Cox_JCP_2015_layers, Fitzner_JACS_2015_HeteroNuc-LJ}

Although we did not explicitly follow nucleation events using local structure measurements, our results will be unaffected by this choice. The drop in energy is typically very sharp, which results in an accurate estimation of the induction time that we use to calculate the nucleation rate constant (with an uncertainty of less than one ns). The differences between induction times of good and bad nucleators are much larger (up to 100 ns difference) and hence results will not  depend strongly on the approach used to obtaining the induction time. This can also be seen in~\cite{Fitzner_JACS_2015_HeteroNuc-LJ} (Figure 2), which shows that the decrease of energy coincides with the increase in ice-like molecules. 

Some key points concerning the reliability of the results are:
\begin{itemize}
 \item The coarse grained mW force field is performing equally well or even better compared to all-atom force fields in many aspects~\cite{Molinero_JPCB_2008_mW}. It is well known however, that the diffusivity is too large in mW, which is one reason why this particular force field is so suitable to simulate the freezing transition. Furthermore, mW has been extensively tested and used in numerous homogeneous and heterogeneous ice nucleation studies~\cite{Moore_PhysChemChemPhys_2010_nanopore, Lupi_JACS_2014_Ccurvessize, Zhang_JCP_2014_surface-structure-INA, Reinhardt_JCP_2014_surface-INA,  Cox_JCP_2015_layers, Cox_JCP_2015_hydrophilicity, Fitzner_JACS_2015_HeteroNuc-LJ, Bi_JPhysChemC_2016_CrystallinityHydrophilicity, Lupo_JChemPhys_2016_interfacialwater-twostep} 
 \item Nucleation rates are converged well enough with 15 nucleation simulations to classify them as good or bad ice nucleators~\cite{Fitzner_JACS_2015_HeteroNuc-LJ}
 \item 4000 water molecules are sufficient to restore the density of bulk water, and enough for finite size effects to play a negligible role in determining whether or not a substrate is a good IN~\cite{Fitzner_JACS_2015_HeteroNuc-LJ}
 \item The lateral dimensions of 50 \AA of simulation box sizes combined with the use of 4000 mW molecules is sufficient, because the critical nucleus size at 205 K for good nucleators is very small (circa 50 molecules)~\cite{Fitzner_JACS_2015_HeteroNuc-LJ}
 \item The qualitative trends observed at a simulation temperature of 205 K remain valid for higher temperatures: good IN at 205 K are also more efficient at nucleating ice at higher temperatures than bad nucleators~\cite{Fitzner_JACS_2015_HeteroNuc-LJ}. The rate constants do change quantitatively, but the distinction between good and bad nucleators still holds at higher temperatures. We focused only on this distinction here, and hence expect our qualitative results to not be affected by the choice of supercooling.
\end{itemize}

\clearpage
\subsection{Structure Search Method}
We used and modified the random structure search approach developed previously~\cite{Pedevilla_JPhysChemC_2016_feldspar-DFT}, in order to calculate how small water clusters adsorb on the all 164 substrates. Details about the approach can be found in~\cite{Pedevilla_JPhysChemC_2016_feldspar-DFT}. To shortly summarize the procedure: Lattices are built based on for example known ice structures such as the basal face of ice I$_\mathrm{h}$. Individual water molecules are then assigned a lattice point on that lattice, and placed onto such a lattice point with an additional random displacement in x, y and z. Furthermore the lattice itself is randomly placed onto the substrate. For each water coverage $n_\mathrm{mW} = \big\{1,2,3,4,5,6\}$, we used a variety of different lattices, namely one corresponding to: the basal face of ice, the primary prism face of ice, the secondary prism face of ice and a uniform rectangular lattice. Furthermore, we also built lattices for coverage $n$ based on structures found via the structure search at coverage $n-1$ for all $n \geq 2$. This is to ensure that stable structures that were previously identified are propagated also for the use at higher coverages. For each substrate and coverage $n_\mathrm{mW}$ we performed a total of 800 structure minimizations (200 for water monomers), leading to a total of more than 650,000 minimization.

For each substrate, we chose the smallest unit cell, but with dimensions of at least 10 \AA~$\times$ 10 \AA~for the structure search. This is akin to simulation cell sizes that are accessible also with \emph{ab initio} methods such as DFT.

\begin{figure}[ht]
\begin{centering}
\centerline{\includegraphics[width=0.8 \textwidth]{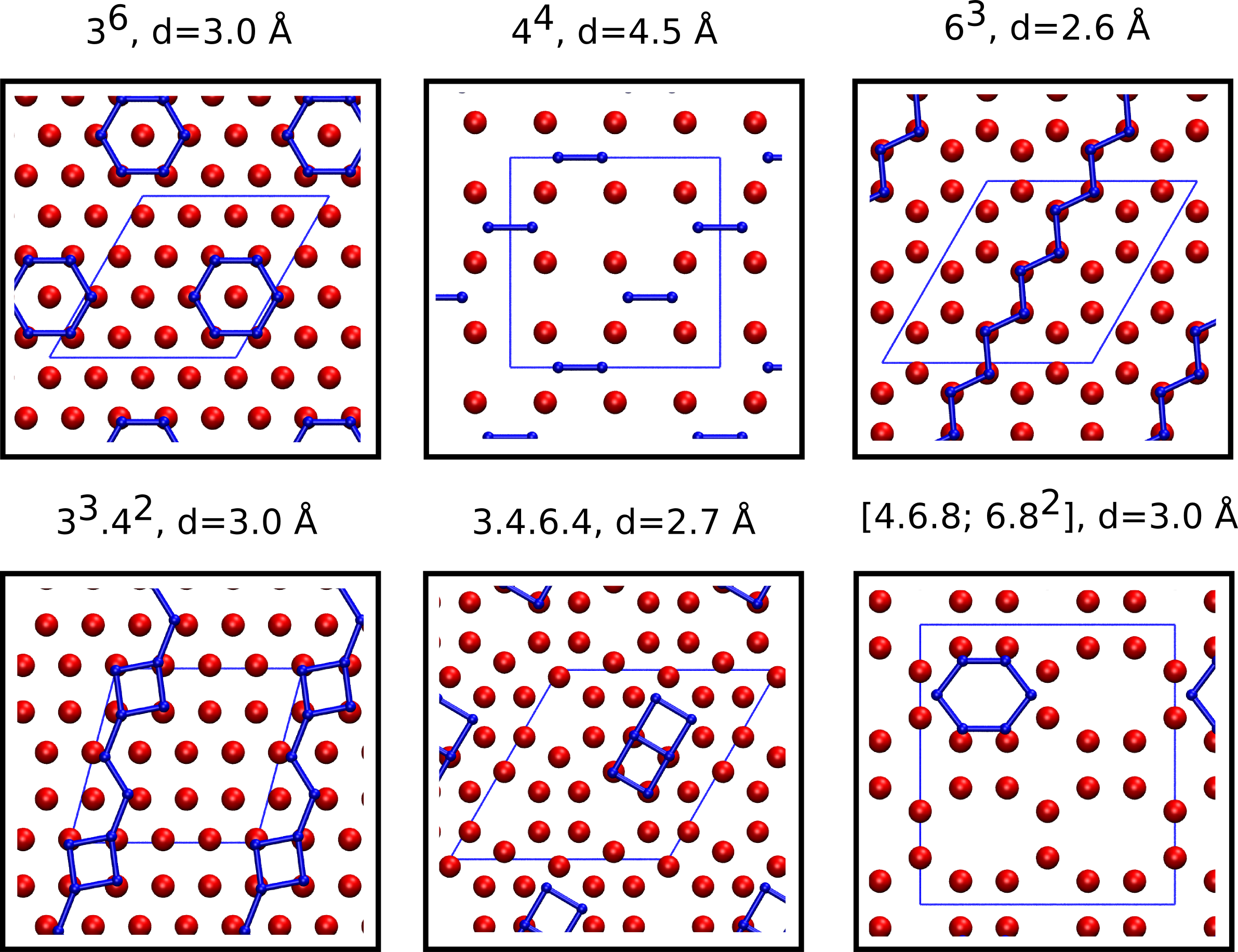}}
\par\end{centering}
\protect\caption{Most stable structures for $n_\mathrm{mW}$ = 6 for six selected substrates to show the structural variety of structures that can form.}            
\label{fig: RSS}
\end{figure}

In figure \ref{fig: RSS} we show few selected structures that we obtained using $n_\mathrm{mW}$ = 6. Whereas on some surfaces, water prefers to cluster together into 1D clusters ($3^6$, $3.4.6.4$, $[4.6.8; 6.8^2]$), on others it prefers to form 2D structures ($6^3$, $3^3.4^2$) or to not cluster together at all ($4^4$).

\clearpage
\subsection{Adsorption Energy Dependence Relation with OH Density}
\begin{figure}[ht]
\begin{centering}
\centerline{\includegraphics[width=0.6 \textwidth]{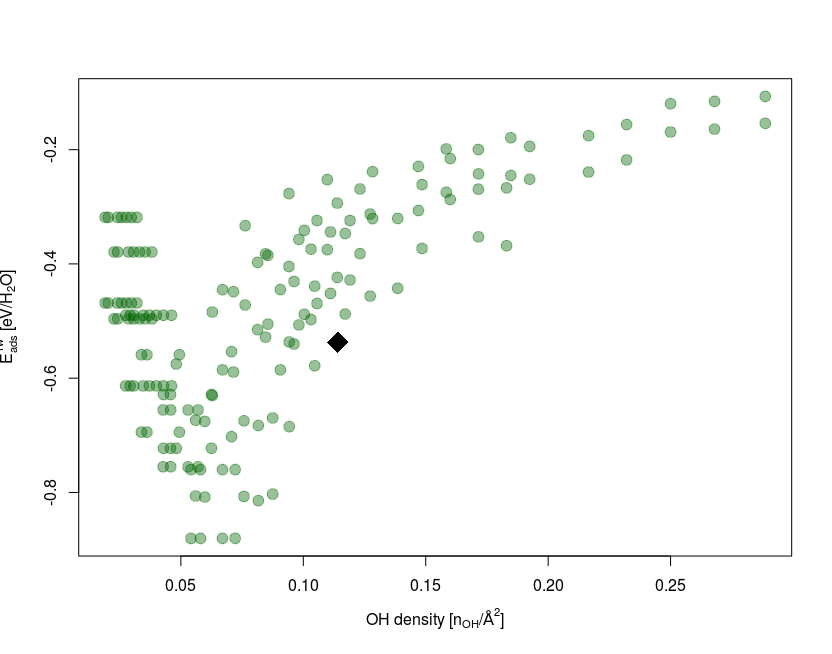}}
\par\end{centering}
\protect\caption{Relation between adsorption energy and OH density. The black diamond shape shows where ice Ih would be in this figure, with the water density shown instead of the OH density.}            
\label{fig: Eads_vs_OHdens}
\end{figure}

In figure \ref{fig: Eads_vs_OHdens} we show the relation between adsorption energy and OH density. Two distinct trends can be identified. At very low OH densities (OH density $< 0.05$ n$_{\mathrm{OH}}$/\AA$^2$) the adsorption energy is independent of the OH arrangement (and density). This is so, because OH groups are so sparse that a water molecule interacts with one and only one OH group, the adsorption energy hence does not depend on how the nearby OH groups are arranged. At OH densities close to ice, the adsorption energies vary quite significantly, and hence are very sensitive to the underlying structure. Structures with densities lower than the ice density tend to have larger adsorption energies, because a water molecules can form several very strong hydrogen bonds at the same time, and interact with the LJ potential in the same time. This scenario is akin to what was found also with DFT on for example water adsorption on feldspar~\cite{Pedevilla_JPhysChemC_2016_feldspar-DFT}. At very high OH densities (OH density $> 0.2$ n$_{\mathrm{OH}}$/\AA$^2$), the underlying OH arrangement again does not have a significant impact on the adsorption energy, because the OH distribution so dense that water experiences it as approximately uniform.

\clearpage
\subsection{Convergence of the Decision Boundary}
\begin{figure}[ht]
\begin{centering}
\centerline{\includegraphics[width=0.42 \textwidth]{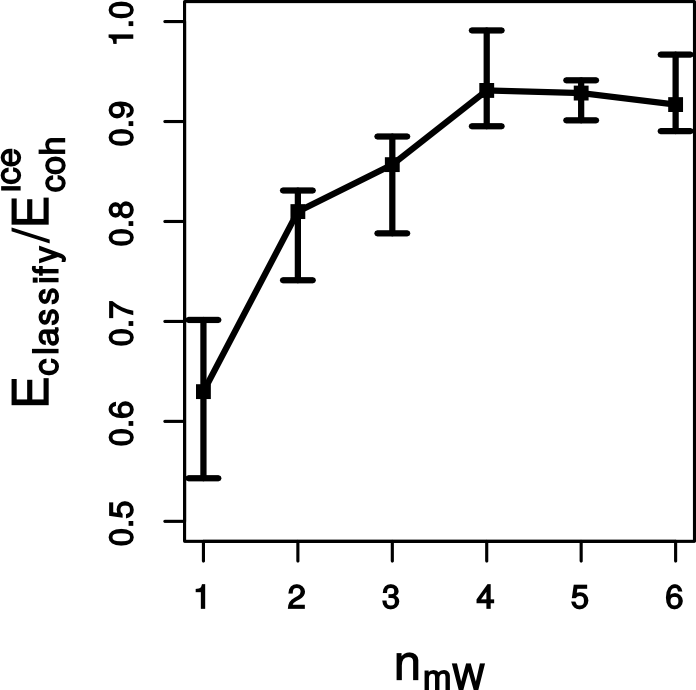}}
\par\end{centering}
\protect\caption{Decision boundary convergence with confidence intervals estimated via a non parametric bootstrap as a function of $n_\mathrm{mW}$.}            
\label{fig: decision boundary convergence}
\end{figure}

The convergence of the decision boundary used to classify good and bad ice nucleators is shown in figure \ref{fig: decision boundary convergence}. In order to account for the fact that we extrapolate conclusions from a finite data set, a non parametric bootstrap was used to estimate confidence intervals for the classification between good and bad ice nucleators~\cite{Efron_CRC_1994_bootstrap}. We resampled our dataset 1000 times (with replacement), and calculated the adsorption energy that classifies between good and bad ice nucleators with maximal accuracy. The confidence intervals are estimated from this using the 2.5 and 97.5 percentile of the resampled distribution. The decision boundary convergences to about 90 \% of $E_{\mathrm{coh}}$, and strictly remains below  90 \% of $E_{\mathrm{coh}}$.

\end{document}